\documentclass[sigconf]{acmart}

\usepackage{booktabs} 

\setcopyright{rightsretained}

\acmDOI{10.475/123_4}

\acmISBN{123-4567-24-567/08/06}

\acmConference[WOODSTOCK'97]{ACM Woodstock conference}{July 1997}{El
  Paso, Texas USA} 
\acmYear{1997}
\copyrightyear{2016}

\acmPrice{15.00}

\begin{document}
\title{A Price Driven Hazard Approach to User Retention}
\titlenote{Produces the permission block, and
  copyright information}

\author{Junhua Chen}
\author{Wei Zeng}
\author{Ge Fan}
\author{Junmin Shao}

\renewcommand{\shortauthors}{B. Trovato et al.}

\begin{abstract}
Customer loyalty is crucial for internet services since retaining users of a service to ensure the staying time of the service is of significance for increasing revenue. It demands the retention of customers to be high enough to meet the needs for yielding profit for the internet servers. Besides, the growing of rich purchasing interaction feedback helps in uncovering the inner mechanism of purchasing intent of the customers.

In this work, we exploit the rich interaction data of user to build a  customers retention evaluation model focusing on the return time of a user to a product. Three aspects, namely the consilience between user and product, the sensitivity of the user to price and the external influence the user might receive, are promoted to effect the purchase intents, which are jointly modeled by a probability model based on Cox's proportional hazard approach. The hazard based model provides benefits in the dynamics in user retention and it can conveniently incorporate covariates in the model. Extensive experiments on real world purchasing data have demonstrated the superiority of the proposed model over state-of-the-art algorithms.
\end{abstract}

%
%



\maketitle

\section{Introduction}

The longer a customer stays on a website, the more loyal the customer was. The length of staying time of users reflects the viscosity of a website and its ability to attract users. Generally speacking, a longer staying time of users increases the chances of commercial advertisements clicks that yield increasing revenue. Frome the user data that Facebook released in April 2016, the average spending time per user on Facebook is 50 minutes more than the combine of that in Instagram and Messenger. The last time Facebook releasing its user time spending data was in July 2014, when users spent more than 40 minutes on Facebook every day. In the corresponding quarter, Facebook's advertising revenue increased by 50\%, the average price of advertising rose 5\%. These data suggested that increasing the user's satying time is significance for increasing the revenue of enterprises.

In order to improve the retention of customers, some businesses tend to collect user feedback to build personalized recommendations services. Others tend to wake up those "dormant" customers by activly sending emails or messenges. Customer retention problem has been extensively studied in many reseach fields, such as telecommunication, financial services and internet services. Among these, the key idea is to distinguish potential churner in the vast sea of people, where churners are defined as those current subscribers who are not likely to renew their subscription in the coming months. Once detected, the churner population is targeted with retention strategies like offers, customer solutions and recommendations to win them back. Whereas Internet is a non-contractual business enviroment that most of the services in which is free of charge. As a result, the lack of financial commitments in Internet services undoubtedly increase the challenge of identifying churner users.

While introducing new users is vital to the business, keeping existing users tend to cost less and be more effective. Reseaches indicated that maintaining long-term customer can produce more revenue for the business than maintaining short-term customer \cite{reinartz2000profitability}. Therefore, the study of return time for exsiting customers has great impact on extending the user's onling time, improving customer loyalty and increasing the income of enterprises. Credit to the rapid development of the Internet, more and more user data becomes available currently. Thus we have enough data to study the customer lifetime duration.

By treating each customer behavior (e.g., visitation behaviors and rating behaviors) as a timestamped events, the timestamp events can be modeled by Poisson processes, a stochastic process assuming the duration time between two events following exponential distribution \cite{kingman1993poisson}. Poisson process is suitable for modeling timestamped events while customers' purchasing behaviors are timestamped events. We assume that user visitation events follows Poisson process and the time duration between visitation behavior follows exponential distribution. A proper intensity is then selected to model these timestamped events and the corresponding parameters are inferred from the history events. Then the intensity of Poisson can be utilized in evaluating the expectation of return time. There is an extension of Poisson process in time duration modeling called hazard model, which is a branch of statistics that deals with the time of occurrence of events. Hazard model is used to predict the probability of the occurrence of an event after $t$ units of time. Two functions is critical in hazard model, survival function and hazard function. Reseaches indicated that hazard can exceed traditional Poisson model in user return time evaluation \cite{kapoor2014hazard}.

Furthermore, recent studies suggested that user's behaviors can be also effected by social factor and external factor \cite{myers2012information,rong2015happened}. Social influence describes that user's behaviors (e.g., review, rating) might get effected by its friends. While external influence suggests that user's behaviors can be influenced by external sources like news media. However, few studies explicitly involve the influence of product attributes, like price of products, quality of products and accessibility of products, upon purchase behaviors. Among those attributes, the retail price usually comes in the first place as one's decision making \cite{zeithaml1988consumer,jagannathan2001dynamics}. To illustrate, when one has a intent to purchase a laptop, she might mark the retail price and pull the trigger as it gets a sale promotion. During her consideration, the gap between retial price and her price perception was finally push she to make the purchase \cite{sato2016model}. In earlier researches, it is suggested that price factor can help modifying recommender systems to be more useful in marketing \cite{schafer1999recommender}. During the past decades, price factor has been introduced in recommendation systems, by treating price as item feature/criteria \cite{burke2002hybrid,zanker2009case}, or by embedding it in a collaborative framework or hybrid recommender system.

Further more, in a more specific scenario, like online retailing, the gap between consumer theories from economics and large-scale recommender systems is further narrowed and Consumer Preferences and Price Sensitivities are modeled simultaneously \cite{wan2017modeling}. The mentioned cases indicate that price factor is one critical aspect in effected the purchase decisions. Therefore, to address the price factor in marketing, we tend to model price perception from an individual level, through a more explicit way as it is studied in economics \cite{zeithaml1988consumer,jagannathan2001dynamics,park2003identifying}.

In this paper, we assume that the intensity of a purchase is composed by three aspects, namely the consilience between user and product, the sensitivity of the user to price and the external influence the user might receive. We propose a price driven hazard model to predict return time of users. Following the Cox's proportional hazard framework, we are able to combine three factors that effect the intensity of purchasing. These three factors are base intensity to capture the interaction of user and product, social intensity to collect influence from neighbors and price intensity to model the willingness of a user in spending.

Furthermore, the Cox's proportional approach has better extension for covariance. As in our model, the influence from neighbors are considered as covariance such that the information is transmitted to the intensity through Cox's proportional method while both base intensity and price intensity account for the baseline hazard rate. The proposed hazard model is then inferred by variational method. Finally, the model is examed under two real world datasets. The approach is aim at providing intuitive and referable suggestions for retailer by giving predictions of user return time under different products with its current resources.

The contributions of this paper are summarized as:
\begin{itemize}
\item We propose a novel hazard based approach to model the user return time of purchasing and incorporate the influence of price factor in making purchasing decisions, which allows to recommend the right item to a user considering its spendable resources by inferring the price perception for each users that denotes how much the user is willing to cost for the purchase.
\item We interpret the collected influence from neighbors as covariates for Cox's proportional methods. It is a non-linear hazard model structure that limit the effect of collected influence serve to be a scalar impact on basic hazard intensity.
\item We investigate the performance of our model through several experiments on few real purchased datasets. The experiment results indicate that our proposed model outperformed other methods in return time predictions, under two difference metrics (RMSE and test log-likelihood). And our methods have consistent performance across various types of users.
\end{itemize}

The rest of the paper is organized as follow: In Section 2 we give a brief review of previous works that most related to ours. In Section 3 we give the definition of our problem and then we cover our models in section 4, 5 and 6. Experiments are shown in section 7 and finally, we conclude our work in section 8.

\section{RELATED WORK}
During the past decades, many approaches were proposed for predicting user return time. Due to space limitations, we can only give a brief review on some important research topics that are the closest to our works. In addition, we briefly introduce several related works with respect to price.

\subsection{Return time prediction}
Return time prediction refers to predicting the next moment of revisit a service or the time of re-purchase based on user purchases history. Each purchase behavior can be regarded as a timestamped event, thus it can be modeled by point processes. The key to point processes is to define a proper intensity fucntion for representing the probability of the occurrence of events. And the expectation of the return time can be computed by the intensity function.

Earlier methods used hazard model, from survival analysis, to model the return time of users \cite{gupta1991stochastic}. They introduced a hazard function to indicate the probability of the current purchase and learn the fucntion by maximizing the likelihood of the history purchases. More recent methods turned to Cox's hazard approaches to model the timestamped events. A Cox's hazard framework can easily incorporate different types of covariates in the model \cite{kapoor2014hazard}, which extends the use of basic point process for additional covariates information. Furthermore, Hawkes processes can model the self-exciting phenomenon, that is, the intensity is no longer independent to its previous events, by introducing collected influence from its precursors, where the influence is represented by trigger kernels (usually a time decay function) \cite{hawkes1971spectra}. Nan Du et al applied a low rank Hawkes process to temporal recommendations, specifically, it is a linear combination of a low rank matrix-factorization and summation of kernals that capture the self-exciting effect. I Valera et al applied the Hawkes process based method for product adoption modeling, where the trigger kernels is used to capture social influence besides self-exciting \cite{valera2015modeling}. Moreover, Yichen W et al overcame the linear structure limitation of Hawkes process in considering self-exciting effect, by introducing isotonic regression to build nonlinear intensity funciton \cite{wang2016isotonic}. Additionally, Du et al explored the connection betweem point processes and recurrent neural network, they passes the hidden layer vector to the intensity function of Poisson to model both user return time and next place with point process simultaneously \cite{du2016recurrent}.

Another popular method for return time prediction is through Poisson factorization \cite{gopalan2014bayesian,charlin2015dynamic,gopalan2015scalable}. Different from the mentioned point processes based methods from above, Poisson factorization methods construct the intensity via matrix factorization approaches. Traditional factorization based methods like tensor factorization allows to uncover pair-wise information within multi-dimensional data \cite{karatzoglou2010multiverse,rendle2010pairwise}. That is, the time information can easily be fed as one dimensional in tensor factorization. Such method can be used in evaluating return time \cite{du2015time}. Poisson factorization is an extension for matrix factorization, which inherit the flexibility of matrix factorization, making it easy to incorporate appropriate implicit feedback such as reviews, clicks and purchases \cite{gopalan2014bayesian,charlin2015dynamic,gopalan2015scalable}. Recently, in the paper by Hosseini et al, the authors introduced a general Poisson factorization framework to combine factorization method to learn pair-wise pattern and exciting information with trigger kernel similar to Hawkes process. The framework is also capable of modeling user's temporal behavior by extending the factorization part \cite{hosseini2017recurrent}.

In addition, authors in \cite{myers2012information,rong2015happened} provide a more top-down modeling method to construct the intensity functions. The intuitively modeling approaches distinguish themselves from the formalist in other point process based methods. The paper \cite{myers2012information} jointly modeled internal influence and external influence on the user behaviors, where the internal influence captured opinions from neighbors and the external influence denoted those outside sources like news and media. Thus two intensity functions were introduce for internal and external influence respectively. Similarly, the authors in \cite{rong2015happened} suggest that users behaviors is supposed to driven by three aspects, namely intrinsic influence, social influence and external influence. And each aspect was assigned with a independent Poisson process and then the combination of the three results in the hybrid model for the occurrence of events.

The mentioned approaches so far rely heavily on implcitly modeling the hazard function over event data that usually ignore important explicit features such as price influence on purchasing. In fact, the impact of the price of products has been considered one of the most important factor in purchasing and heavily studied by many works in economics and marketing.

\subsection{Price and price sensitivity}
In the past few years, the retial price of the product has been widely concerned and studied. On one hand, retial price of the product as one of the important features of the product, is applied to various recommendation systems to yield more accuracy recommendation algorithms \cite{schafer1999recommender,burke2002hybrid,zanker2009case,montgomery2009prospects,sato2016model,guo2017recommend,wan2017modeling}.

Price factor has been promoted in many personalization and recommender systems. Most of them are feature-based methods, i.e. price factor is regarded as additional item features in recommendation systems. Then the price-based features were handled by context-aware framework or hybrid recommendation systems \cite{burke2002hybrid,zanker2009case}. Some indirect features induced by price such as discount rate was also considered in recommendation via a collaborative filtering framework \cite{sato2016model}. Meanwhile, it is also reasonable to interpret price factor as user features. Authors In the paper \cite{guo2017recommend} model price preference for each users through fuzzy set theory and collective filtering. Further more, another price induce feature, price sensitivity which indicate how the choice of user is influenced by the price, has been studied in \cite{wan2017modeling}. The paper concludes that both user preferences and price sensitivities are critical in online shopping. In addition, user's capacity can effect its next purchase in online shopping \cite{kooti2016portrait}. Price of trust is quantified in paper \cite{guo2011role} to denote the amount of money will a consumer pay for transaction with who she/he trust. In these works, price can be served as direct features or price can induce new features to better describe the data.

On the other hand, the retail price of the product is deemed as one of the product quality evaluation criteria, profoundly affecting the user purchase behavior \cite{zeithaml1988consumer,lichtenstein1993price,grewal1998effect,jagannathan2001dynamics,park2003identifying}.

In the marketing literature, price factor is considered a fundamental feature of product. Hypothesis upon how price effect consumer in its behaviors are build, to illustrate, retial price can in some way suggest the quality of the product, mirror the cost of the purchase and hint the value of gain by the purchase \cite{zeithaml1988consumer,grewal1998effect}. Pricing is critical in effecting the polarity of the purchase intention for customer \cite{lichtenstein1993price}. Too high the price might negative the sales and too low the price reduces potential revenue. Besides direcly study the relationship between price factor and other variables, the relationship of which can be builded in a obliquely way. For instance, price factor is closely related to revenue, authors in \cite{jagannathan2001dynamics} ratified their relationship based on empirical studies upon customer purchasing power and their willingness to pay. In the paper \cite{park2003identifying}, the authors believed that price factor, belonged to features of the product and the retialer iteself, is just one of the many factors that can effect the purchase decision of customer. These papers are aim at providing insides in building price based model. The concepts and hypothesis within are usually tested by questionnaire study.

In summary, previous approaches on predicting return time is mainly focus on constructing the a proper intensity function that can better describe the probability of event occurrence under specified scenario. However, in online shopping, price factor is critical in many ways yet it seldom makes the way into point process studies. Inspired by attention of price factor for modeling user behavior in marketing, we propose a novel point process approach driven by price to evaluate user retention in online shopping.

\section{Survival Analysis}
Point process is widely used to model visitation behavior of a user on free web services in recent years. For instance, user-item interaction events can be modeled to generate from a Poisson process such that the duration times between events follow an exponential distribution. Generally speaking, an interaction event $e_i$ can be represented by a triple $(u_i,o_i,t_i)$, which indicats that user $u_i$ purchased product $o_i$ at time $t_i$. Let  $ \mathcal{H}(t_N)=\{e_i \}^{M(t_N)}_{i=1}$ denotes the set of user-item interaction events until time $t_N$ where $M(t_N)$ is the number of interactions up to time $t_N$. By exploiting user-item interactions data before time $t_N$, one can predict the possible occurrence of user-item interactions in the future. Poisson process is one of the most popular methods in modeling events sequence. The main idea behind Poisson process is to define an intensity function to describe the occurrence of events \cite{kingman1993poisson}. Survival analysis is an extension of Poisson process that focuses on modeling time duration that has a wide range of applications in economics, medicine, engineering and sociology. \cite{gupta1991stochastic,kapoor2014hazard}. Survival analysis offers a rich set of methods that allow us to easily address questions like what is the probability of an event occurring after $t$ units of time or what is the future rate of occurrence of the event given it has not happened in $t$ units of time.

Two functions are critical for analyzing duration time between events.

\textbf{A survival function} to describe the probability of an event $e_{i}$ still "live" for $t$ units of time, i.e., the event has not happened yet with the elpased $t$ time.
\begin{equation}
\label{sa_prob}
S_{i}(t) = P(T > t),
\end{equation}
where $T$ is a random variable to denote elapsed time or duration time.

\textbf{A hazard function} to measure the instantaneous "death" rate of an event $e_{i}$, i.e., the probability of the event occurring after $t$ units of time conditioned on it has not happened yet.
\begin{equation}
\label{sa_lambda}
\lambda_{i}(t) = \lim_{dt \to 0} P(T < t+dt | T \ge d ) = \frac{-S_{i}^{'}(t)}{S_{i}(t)}.
\end{equation}
It is worth noting that hazard function is neither a probability nor density as it does not integrate to one.

One can express the survival function in terms of hazard fucntion or vice versa with Eq\ref{sa_prob} and Eq\ref{sa_lambda}:
\begin{equation}
\label{sa_s}
S_{i}(t)=e^{-\Lambda_{i}(t)}=e^{\int^{t}_{0}{ \lambda_{i}(\tau)d\tau }},
\end{equation}
where the $\Lambda_{i}(t)$ is defined as cumulative hazard.

In practice, hazard function $\lambda(t)$ is exclusively designed for real-world applications. For instance, in music recommendation service, in order to consider covariates like user activities including number of visits per week, replay times of a song or number of revisit to a artist, a Cox's proportional hazard is up to the task of building a model with covariates. In Cox's proportional hazard model, the covariates can only effect the magnitude of hazard rate and leave the shape of the hazard function untouched \cite{kapoor2014hazard}. An other example, to consider how the history listen behavior will influence the future, one can add self-exciting mechanism in the hazard rate \cite{du2015time}. Overall, an exclusively hazard function defines the intensity of an event occurring after $t$ units of time conditioned on the elpased $t$ time. With a defined hazard function as in Eq\ref{sa_lambda}, the likelihood of  events in observed time period $[0,t_N]$ can be expressed as
\begin{equation}
\label{sa_like}
\mathcal{L}(\mathcal{H}(t_N))=\prod_{e_{i} \in \mathcal{H}(t_N)}{ \lambda_i(t)  S_i(t)} ,
\end{equation}
where the front component including $\lambda_i(t)$ represents the likelihood of all occurred events and the rear component of $S_i(t)$ contains the likelihood of censored data, i.e., no event has occurred in those time units. It is worth noting that superposition property also hold in hazard model, which makes it fairly easy for further extension. Being identical with Poisson process where the superposition of independent Poisson processes is a Poisson process with the intensity that is the summation of intensities of those independent ones, the sum of two hazard rates corresponds to a superposition of two Poisson processes.

\section{Price Perception Model}
As mentioned earlier, price has been considered as a critical factor in influencing customer decisions and has been widely studied in economics and psychology \cite{puccinelli2009customer}. Nevertheless, the mainstream methods for duration time modeling such as Poisson processes and hazard model are still lack of attention to explicit price modeling.

In this paper, we assume that the purchase intent of a user to a product is steered by its price perception factor along with the collective influence from its neighbors. For example, a user in Amazon might make a purchase mainly due to her interest to the product. But her budget at that time would significantly effect her final decision. In our model, the user price perception factor is served along with the base intensity that tuned by the covariates which is the collected neighbors influence. Following the Cox's proportional hazard model, an exponential activation function is adopted to transfer the covariates which can only effect the magnitude of the hazard rate.

In the purchase events scenario, each purchase event $e_{i}$ is thus contributed by the three intensity components, namely the intensity of price perception per user $\lambda^{price}_{i}$, the base intensity to cross-link each user and product $\lambda^{base}_{i}$ and the social intensity $\lambda^{social}_{i}$ that collects the neighbors influence of user $u_i$ with respect to product $o_i$. It can be expressed mathematically as:

\begin{equation}
\label{pp_lambda}
\lambda_{i} = \lambda^{base}_{i} \lambda^{price}_{i} \lambda^{social}_{i}.
\end{equation}

\begin{figure}
\includegraphics[height=2.3in, width=2.3in]{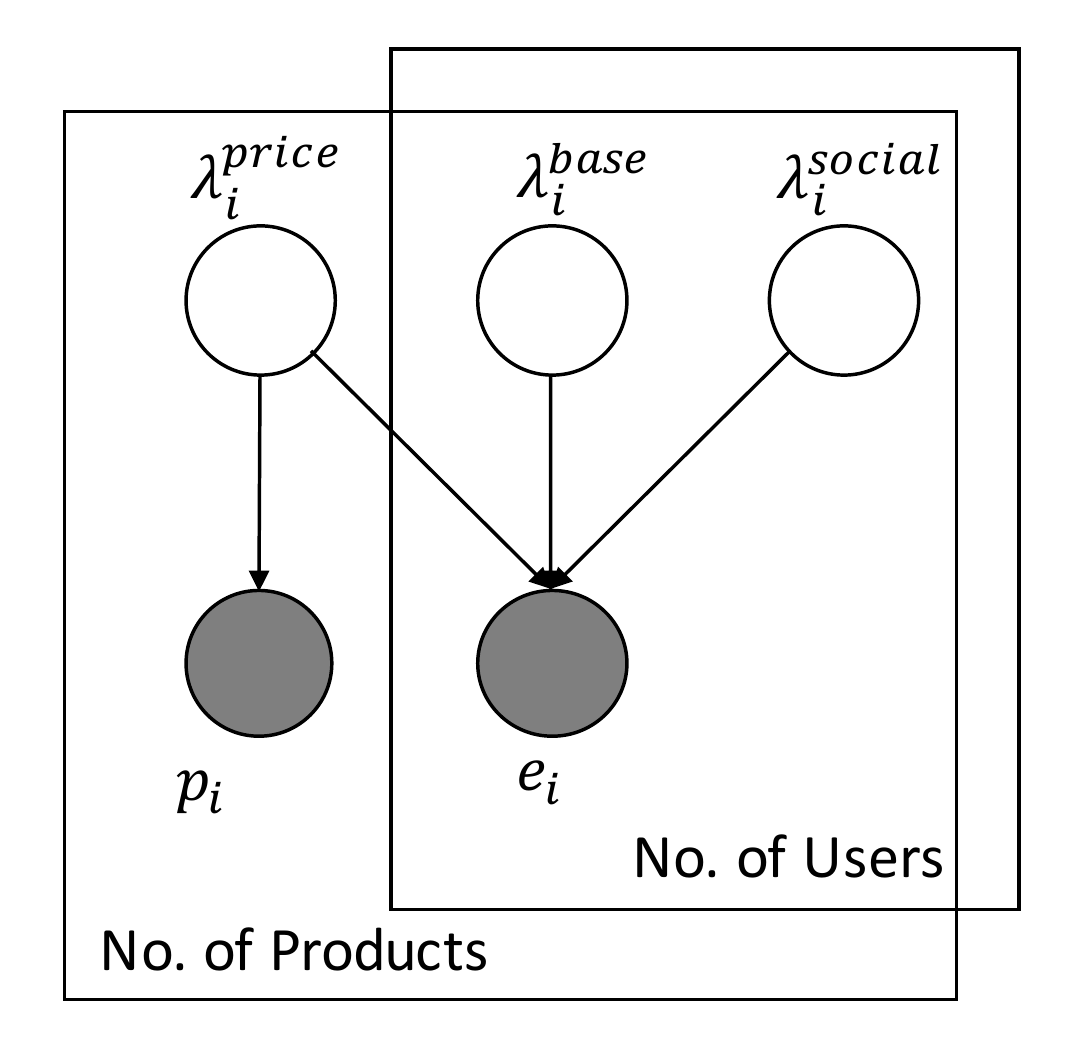}
\caption{Illustration of our proposed mixed hazard model. Three intensity components of event $e_i$, namely $\lambda^{price}_{i}$, $\lambda^{base}_{i}$ and $\lambda^{social}_{i}$, are contributed for the generating of purchase event $e_{ui}$. In addition, the price intensity $\lambda^{price}_{i}$ also accounts for the generating of all the spending $\mathcal{P}_{u_i}$ of user $u_i$.}
\label{Illu}
\end{figure}

In addition, the data of spending of each user is collected for further modeling the user price perception. An illustration of the proposed model is presented in figure \ref{Illu}.

A common choice for the base intensity $\lambda^{base}_{i}$ is to assume this basic intensity to be independent of time. For instance, the most naive case where $\lambda^{base}_{i} = \lambda_0$ such that $\lambda_0$ is a globle parameter. But this naive model fail to model the interaction of user-product and fail to capture the influence of the characters (i.e., user and product) of this event. In order to build a more personalized intensity. The base intensity is assumed to capture the preference of user $u_i$ to product $o_i$ with the following equation.
\begin{equation}
\label{lambda_base}
\lambda^{base}_{i} = \theta_{u_i}  \theta_{o_i} ,
\end{equation}
where $\theta_{u_i}$ is the latent factor of user $u_i$ and $\theta_{o_i}$ the latent factor of product $o_i$. In our model, we choose the $\theta$ as number and the base intensity as their product. However, our framework allows for more complex expansion for the base intensity.

In the following, we introduce how to model user price perception and influence from neighbors. Then we show how to mixed these factors into a joint Cox's proportional hazard model.

\subsection{Modeling}
\textbf{Modeling User Price Perception.} Multiple researches indicate the willingness to purchase is closely related to the price of the product \cite{zeithaml1988consumer,lichtenstein1993price,grewal1998effect}. In addition, considering the privilege of Gamma distribution in modeling globel behaviors in early research \cite{gupta1991stochastic}, we choose Gamma distribution to model the bidding distribution of users.

Furthermore, noting the user's bid is related to her purchase frequence \cite{kooti2016portrait}. We parametered the Gamma distribution with shape parameter equals to the cumulated purchases of user $u$, $\mathcal{N}^{t}_{u}$ and with rate parameter equals to user price perception $\kappa_u$. The bidding distribution can be written as:
\begin{equation}
\label{pdf_price}
f(p;\mathcal{N}^{t}_{u},\kappa_{u})=\frac{ \kappa_{u}^{\mathcal{N}^{t}_{u}} p^{{\mathcal{N}^{t}_{u}}-1} e^{-\kappa_{u}p} }{ \Gamma(\mathcal{N}^{t}_{u}) },
\end{equation}
where $p$ is the bidding price and the user price perception $\kappa_u$ is drawed from a Gamma distribution to denote the price intensity. Samller $\kappa_u$ induce smoother Gamma density that it is easier to draw high bid from. Larger $\kappa_u$ indicates the user's bidding pattern is more consist with common people. In our model, we let the price intensity to direcly effect the base intensity. By multiplying base intensity and price intensity, a simple homogeneous hazard model can be written as:
\begin{equation}
\label{lambda_baseprice}
\lambda^{base+price}_{i}= \underbrace{\theta_{u_i} \theta_{o_i}  }_{base} \underbrace{ \kappa_{u_i} }_{price} .
\end{equation}

\textbf{Modeling User Neighbors Influence.} It is common that the occurrence of an event can be influenced by its precursors. A precursor means an event happened before the event of concerned. For example, one might take recommendation from her friends who is similar to her interest \cite{rong2015happened}. Or one might hesitate after seeing the reviews of others before making the purchase on a product. These behaviors can be concluded as collected influence from precursors. In our model, the precursors of an event $e_{i}$ is defined as $E_{i}=\{ e_{j}|t_j < t_i  \}$. We assume that any events happened before the objective event $e_{i}$ can lay a influence upon it. In our scenario, any users who has purchased product $o_i$ can have influence on the occurrence of $e_{i}$. Besides, the influence decreases by time. In practice, an exponential decay function $\gamma(\Delta t;\sigma)=e^{- \sigma \Delta t}$ is adopted to simulate the real world influence decrease, where $\Delta t$ denotes the elpased time between the precursors and our objective event. The larger the elpased time, the lighter the influence. Following the diffusion model, we assume the influence of precursors is diffused from user to user within a underlying latent network \cite{myers2012information,rong2015happened}. Thus, each user is assigned with an influence rate $\alpha$ measuring how likely a user is to influence others and an infection rate $\beta$ measuring how likely a user will get effected by others. The descriptions above can be formulized as
\begin{equation}
\label{lambda_social}
\lambda^{social}_{i}=\sum_{e_{j} \in E_{i}}{ \alpha_{u_j} \beta_{u_i} \gamma(\Delta t)}.
\end{equation}

\textbf{Mixed Hazard Model.} In our model, the hazard function consists three parts, namely base intensity, price intensity and social intensity. These three intensities are mixed through Cox's proportional hazard method. Cox's model follows a basic assumption that the covariates can only effect the magnitude of the hazard rate without effecting its shape. In our model, the covariates is the collected neighbors influence as shown in Eq\ref{lambda_social}. and the baseline hazard rate is the combination of base intensity and price intensity as shown in Eq\ref{lambda_baseprice}. The hazard function of our model can be written as follow：
\begin{equation}
\label{lambda_mix}
\lambda^{\ast}_{i}(t)=\underbrace{\theta_{u_i}  \theta_{o_i}  }_{base} \underbrace{ \kappa_{u_i}}_{price} \underbrace{ e^{  w(t)\sum_{e_{j} \in E_{i}}{ \alpha_{u_j} \beta_{u_i} \gamma(\Delta t)}
  }}_{social} .
\end{equation}

We call it Mixed Hazard Model (MHM). Roughly speaking, the front component is represented by base intensity that captures the interaction of user and product \cite{hosseini2017recurrent}. Middle component is price intensity, $\kappa$, that account for user price perception by steering the user's bidding Gamma distribution. The rear component is social intensity storing the collected influence from neighbors. The $w(t)$ is the parameters for covariates since the collected influence is deemed as covariates in our model and $w(t)$ is drawed from Gaussian random walk. With the definition of hazard function, one can yield its survival function with Eq\ref{sa_s},
\begin{equation}
\label{sur_mix}
S^{\ast}_{i}(t) = e^{-\Lambda^{\ast}_{i}(t)},
\end{equation}
where $\Lambda^{\ast}_{i}(t) = \sum^{t}_{\tau=0}{ \lambda^{\ast}_{i}(\tau) }$ is the cumulative hazard function.
Overall, the likelihood of the proposed model is
\begin{equation}
\label{like_mix}
\mathcal{L}\left(\mathcal{H}(t),\mathcal{P},\Theta \right)=\prod_{e_{i} \in \mathcal{H}(t),p \in \mathcal{P}}{ \lambda_{i}^{\ast}(t_i) S_{i}^{\ast}(t_i)} f(p_i;\mathcal{N}^{t}_{u_i},\kappa_{u_i}) ,
\end{equation}
where $\Theta$ denotes the set of all parameters in the model and $\mathcal{P}=\{\text{price }p_i \text{ of all events } e_{i} \in \mathcal{H}(t)\}$ is the set of price of all purchasing happened in $\mathcal{H}(t)$.

\begin{table*}
\centering
\begin{tabular}{lccccc}
\toprule
Dataset       & No. of Users & No. of Items & No. of Purchases & Duration & Started Time \\ \midrule
Video Game    & 7,724        & 15,307       & 137,323          & 15                     & 01/2013      \\ 
Food          & 6,280        & 15,841       & 96,160           & 15                     & 01/2013      \\ 
Movie \& TV   & 44,187       & 49,624       & 1,164,219        & 15                     & 01/2013      \\ 
Online Retail & 3,756        & 2,882        & 391,773          & 12                     & 12/2010      \\ \bottomrule
\end{tabular}
\caption{Statistic of various categories products datasets in Amazon and Online Retail. Amazon dataset (Video Game, Food and Movie \& TV) contains 15 months purchases log started from 01/2013. 12 months are selected for inferring the model and evaluated on the rest 3 months. Online Retail data contains 12 months transactions started from 12/2010 and 9 months are used for training and the rest 3 months for evaluation. In both Amazon and Online Retail data, we filter users and items whose degree is less than 10.}
\label{Dataset}
\end{table*}

\subsection{Predicting}
After infering all model parameters $\hat{\Theta}$, the hazard rate for a future purchase event $e^{\ast}_{i}$ can be written as:
\begin{equation}
\label{lambda_pred}
\hat{\lambda}^{\ast}_{i}(t|\mathcal{H}(t),\mathcal{P},\hat{\Theta}) =  \hat{ \theta}_{u_i} \hat{\theta}_{o_i}  \hat{\kappa}_{u_i}  e^{  w(t)\sum_{e_{j} \in E_{i}}{ \hat{\alpha}_{u_j} \hat{\beta}_{u_i} \gamma(\Delta t)}  }.
\end{equation}

Then the cumulative hazard function $\hat{\Lambda}^{\ast}_{i}(t)$ and survial fucnton $\hat{S}^{\ast}_{i}(t)$ can be computed accordingly. The expected time of return for the future purchase event $e^{\ast}_{i}$ can be computed by the equation below as in \cite{kapoor2014hazard},
\begin{equation}
\label{exp_pred1}
E(T)=\int^{t_d}_{0}{ \hat{S}^{\ast}_{i}(\tau)d\tau },
\end{equation}
where $t_d$ is the time period of concern, i.e., $t_d=60$ denotes we concern the purchases might occurr within the next 60 days. In practice, we replace the integrating in Eq\ref{exp_pred1} with summation which yields a new expected time written as
\begin{equation}
\label{exp_pred2}
E(T)=\sum^{t_d}_{\tau=0}{ \hat{S}^{\ast}_{i}(\tau)d\tau }.
\end{equation}

\subsection{Inference}
The model can be learned from maximizing the joint likelihood in Eq\ref{like_mix}  providing the history purchases $\mathcal{H}(t)$ and price information $\mathcal{P}$. For convenience, we denote all the parameters in our model as $\Theta=\{ \theta, \kappa, \alpha, \beta, h\}$, where $h$ stands for all hyperparameters for the priors considered. We exploit the mean field automatic differentiation variational inference (ADVI) \cite{kucukelbir2017automatic} to learn the parameters of our model. Variational inference methods approximate the posterior by defining a variational family of distributions over the hidden variables and then find a distributions that is close to the true posterior by minimizing the Kullback-Leibler divergence between them. ADVI is a extension of variational inference methods that it chooses approximative distributions for posterior only from Gaussian by first reparameterize the original distributions \cite{kucukelbir2017automatic}.

The mean field distributions are considered to be independent from each other which yields the factorization as
\begin{equation}
\label{vi_fact}
q\left( \theta, \alpha, \beta, \kappa, w\right) =
\prod_{u=1}^{M}{ q(\theta_u) q(\alpha_u)  q(\beta_u) q(\kappa_u)}
\prod_{i=1}^{N}{ q(\theta_i) }
\prod_{t=0}^{t_N}{ q(w(t)) }.
\end{equation}

Applying the ADVI algorithm, each variational distribution is chosen from Gaussian distribution. But to ensure the inference work correctly, it is required to reparameterize the original distributions. In our model, $w$ is sampled from a Gaussian distribution and $\kappa$ is sampled from a Gamma distribution. The rest variables have uniform prior. Thus, a log transformation is performed on variables $\kappa$ and the rest variables use logit transformation. Let $\mathcal{T}$ denote the transformation and $z$ denote the hidden variable. By the mean field variational method, the optimal approximate functions for each variable can be written as
\begin{equation}
\label{vi_infer}
\text{ln}q^{\ast}_{j} = \mathbb{E}_{i \neq j}\left[ \text{ln}p\left(\mathcal{T}^{-1}\left(z\right);\mathcal{H}(t),\mathcal{P},h\right) + \text{ln}\left| \text{det}J_{T^{-1}}\left(z\right) \right| \right] + \text{const}.
\end{equation}

The gradient of the Eq\ref{vi_infer} is then approximate by Monte Carlo method. After acquiring  the gradient, Adagrad is used for optimization \cite{duchi2011adaptive}. Our model is build based on the tookit PyMC3 \cite{salvatier2016probabilistic} which is a probabilistic programming tool written in python.

\section{Experiments}
In order to evaluate the performance of our proposed MHM model, we conduct experiments on two large real world datasets, Amazon purchased datasets and online retail transaction data. We demonstrate the effectiveness of MHM in duration time prediction. Further experiments acorss different users and events show the superiority and robustness of the propsoed model over other comparisons methods.

\subsection{Datasets}
\setcounter{footnote}{0}

We evaluate the predictive performance of our method on two real world datasets from different domains.

\textbf{Amazon Purchased Dataset.} Amazon is one of the largest online retialer in the world. In our experiments, we analysis three Amazon products categories, namely Video Game, Food and Movie \& TV\footnote{http://jmcauley.ucsd.edu/data/amazon/links.html} \cite{he2016ups}. More detialed statistic of the data can be found in Table \ref{Dataset}. We collect purchases log from January 2013 till April in the next year. Each purchase in Amazon datasets can be represented as a tuple $e_i = (u_i,o_i,t_i)$. In addition, we include the price paid in each purchase data to infer the bidding pattern of users.

\textbf{Online Retail Transnational Data.} This transnational dataset contains transactions for a UK-based and registered non-store online retail\footnote{https://archive.ics.uci.edu/ml/datasets/Online+Retail} \cite{chen2012data} with detials shown in Table \ref{Dataset}. The company mainly sells unique all-occasion gifts. Many customers of the company are wholesalers. We collect a year transactions in 2011. The transactions data includes customers ID, StockCode as products ID along with its price and the transactions time. For convenience, we denote these two datasets as Amazon and Online Retial. Noting that we exclude users and products who appearing less than 10 times in the datasets.

\subsection{Baselines}
To evaluate the predictive performance of forecasting user return time, we compare with the following point process models:

\textbf{Poisson Process (PP).} It is a relaxation of our model that only take consider the base intensity in modeling occurrence of events as in Eq\ref{lambda_base}. In this baseline, the intensity is a constant for its corresponding user and item, regardless of the history purchases.

\textbf{Influence Based (IB).} This baseline captures the influence diffusion with the latent user network. Each user is assigned with specific influence rate and effected rate. The intensity function follows Eq\ref{lambda_social}. By linearly adding the collective neighbors influence to the base intensity of Poisson process, we got a linear combination model of the base intensity and social factor that effect the user purchase intent. This model is denoted as Influence Based in our comparisons.

\textbf{Cox's Combination (CC).} Cox's proportional hazard model provides us a strong method to consider base intensity and covariance. By treating the collective neighbors influence as covariance, we combine the base intensity and social intensity by Cox's proportional hazard framework \cite{kapoor2014hazard}. Note that the major difference between this model and our proposed model is whether including the price intensity or not. Thus, we also denote this baseline as without price approach to specifically examine the legality of considering price in our model.

\textbf{Hawkes Process (HP).} In terms of formula form, Hawkes process is very similar to the previously mentioned baseline, Influence Based. They share the same linear structure and same base intensity. However, the difference between these two baselines is that Hawkes process focuses on modeling self-exciting information, that is, the collective influence of history purchases for the users in our problem. While Influence Based aims at capturing the collective influence of neighbors which is the influence from who purchased the same product.

\begin{table*}[htbp]
\centering
\begin{tabular}{rlllllllllll}
\toprule
\multicolumn{1}{c}{}      & \multicolumn{5}{c}{Video Game}        &  & \multicolumn{5}{c}{Food}             \\ \cline{2-6} \cline{8-12} 
\multicolumn{1}{l}{}      & 30    & 45    & 60    & 75    & 90    &  & 30   & 45    & 60    & 75    & 90    \\ \midrule
\multicolumn{1}{l}{Model} &       &       &       &       &       &  &      &       &       &       &       \\
PP                        & 9.46  & 15.19 & 21.46 & 28.26 & 34.20 &  & 9.95 & 15.44 & 21.10 & 27.08 & 33.11 \\
HP                        & 9.35  & 14.73 & 20.61 & 27.00 & 32.80 &  & 9.89 & 15.34 & 20.97 & 27.04 & 33.21 \\
CC                        & 9.48  & 15.25 & 21.55 & 28.28 & 34.09 &  & 9.97 & 15.48 & 21.19 & 27.13 & 33.17 \\
IB                        & 9.47  & 15.21 & 21.41 & 28.09 & 34.00 &  & 9.94 & 15.42 & 21.12 & 27.16 & 33.24 \\
MHMl                      & 9.05  & 13.55 & 18.24 & 23.43 & 27.74 &  & 9.26 & 13.64 & 18.03 & 22.56 & 27.41 \\
MHMe                      & 9.05  & 13.53 & 18.23 & 23.34 & 27.41 &  & 9.22 & 13.58 & 17.93 & 22.30 & 26.99 \\ \bottomrule
\toprule
\multicolumn{1}{l}{}      & \multicolumn{5}{c}{Movie \& TV}       &  & \multicolumn{5}{c}{Online Retial}    \\ \cline{2-6} \cline{8-12} 
\multicolumn{1}{l}{}      & 30    & 45    & 60    & 75    & 90    &  & 30   & 45    & 60    & 75    & 90    \\ \midrule
\multicolumn{1}{l}{Model} &       &       &       &       &       &  &      &       &       &       &       \\
PP                        & 10.10 & 15.61 & 21.55 & 27.81 & 34.24 &  & 9.80 & 14.34 & 21.35 & 28.68 & 36.59 \\
HP                        & 9.69  & 14.74 & 20.19 & 26.25 & 32.48 &  & 9.51 & 14.38 & 21.85 & 31.94 & 42.56 \\
CC                        & 10.13 & 15.69 & 21.71 & 28.02 & 34.50 &  & 9.63 & 14.05 & 21.15 & 28.03 & 35.62 \\
IB                        & 10.09 & 15.58 & 21.48 & 27.71 & 34.11 &  & 9.65 & 14.07 & 21.05 & 27.88 & 35.50 \\
MHMl                      & 9.14  & 13.90 & 18.70 & 23.98 & 29.37 &  & 9.12 & 13.27 & 18.92 & 25.53 & 33.21 \\
MHMe                      & 9.15  & 13.89 & 18.67 & 23.94 & 29.34 &  & 9.13 & 13.31 & 18.92 & 25.56 & 33.09 \\ \bottomrule
\end{tabular}
\caption{Predictive performance on RMSE. Four tables are corresponded to four different datasets, i.e., Video Games, Gourmet Food, Movie \& TV and Online Retail. The tables show the RMSE results for duration time prediction under different deadline.}
\label{Res_RMSE}
\end{table*}

\begin{figure*}[htbp]
\centering
\includegraphics[height=2.12in, width=6in]{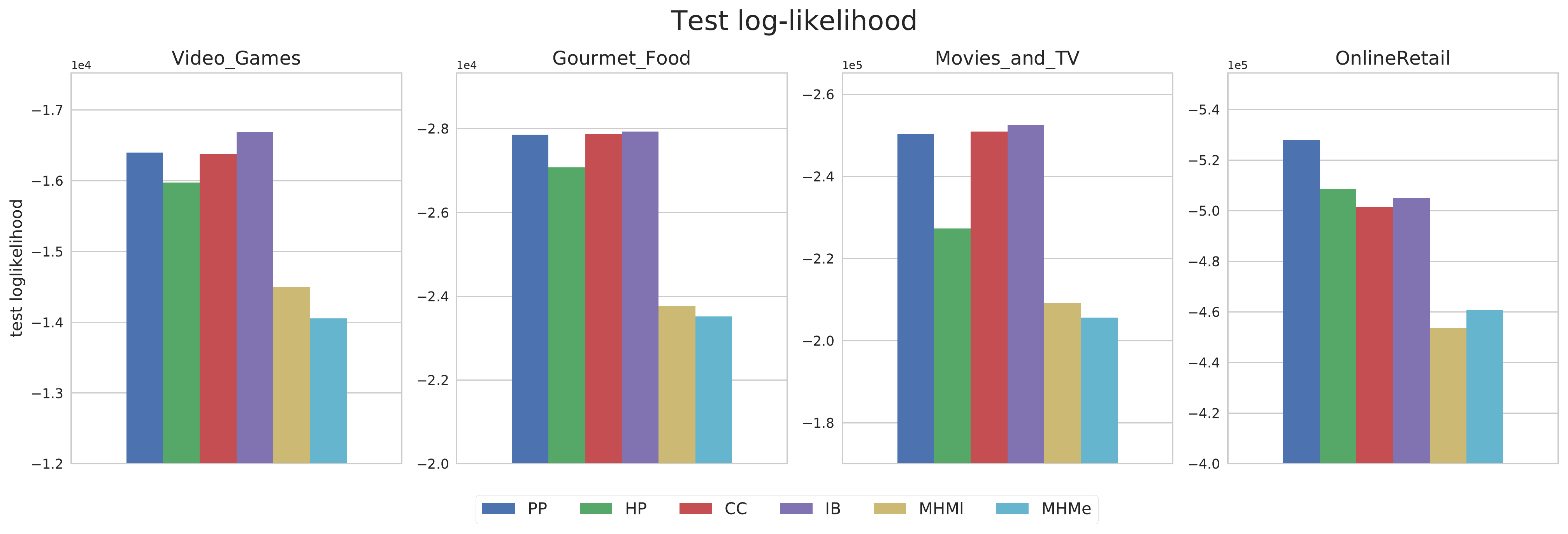}
\caption{Predictive Performance on log-likelihood on test set. The subplots correspond to test log-likelihood on various dataset (larger the better). Different approaches are denoted by different bars. Our proposed models outperform other methods on all datasets by a considerable margin.}
\label{Res_TLL}
\end{figure*}

\subsection{Performance Evaluation}
We performed serveral experiments on real world datasets to examine the effectiveness of MHM model. Noting that in our comparisons, there are linear and exponential structure, thus in order to better elucidate the performance of MHM. We included two various of MHM model. One has exponential structure  denoted as MHMe. MHMe is expressed in Eq\ref{lambda_mix}. The other has linear structure similar to one of the comparisons IB. It can be written as
\begin{equation}
\label{lambda_linear}
\lambda^{\ast}_{i}(t)= \theta_{u_i}\theta_{o_i}  \kappa_{u_i}  + {  \sum_{e_{j} \in E_{i}}{ \alpha_{u_j} \beta_{u_i} \gamma(\Delta t)}  } .
\end{equation}

We denote this various of MHM as MHMl. In our experiments, we focus on evaluating our model performance with return time prediction task. In addtion, effectiveness of model is also examined on likelihood task.

\textbf{Experimental Setup.}

In this section, we show the setup for the return time prediction task and likelihood task. For return time prediction, we split the Amazon and Online Retail dataset according to time. In detail, purchases within 2013 constitute our training set and purchases in the next three months constitute our testing set. In Online Retail dataset, the training set contains transactions in nine months started from January 2010 and transactions in the last three months constitute the testing set. The training set is denoted as $\mathcal{H}=\{ \mathcal{H}(t_N), \mathcal{P}\}$ and the training set is denoted as $\mathcal{T}=\{ \mathcal{H}(>t_N),\mathcal{P} \}$. The return time can be evaluate by the expectation of survival time with Eq\ref{exp_pred2}. Since hazard model or more generally, Poisson processes, can be defined with the hazard function or intensity rate, we utilized Eq\ref{exp_pred2} as our time prediction function for MHM and all its comparisons. Noting that in Eq\ref{exp_pred2}, there is a $t_d$ indicating the time period of concern. In practice, we enumerate $t_d$ every other 15 days from 30 to 90 days. The metrics for evaluating return time prediction is RMSE:
\begin{equation}
\label{rmse}
RMSE = \sqrt{ \frac{1}{|\mathcal{T}|} \sum_{e_i \in \mathcal{T}}{ (t_{i} - \hat{t}_{i})^2 } }.
\end{equation}

Furthermore, we also evaluated the predictive performance of each model based on the log-likelihood for the testing set \cite{kim2014tracking}. The test log-likelihood (TLL) is computed as follows,
\begin{equation}
\label{tll}
TLL = \sum_{e_i \in \mathcal{T}}{\text{log}p(e_i  | \hat{\Theta} )}.
\end{equation}

\subsection{Experimental Results}
In the following, we show the predictions results of return time and the results of test log-likelihood on various datasets. And we further analysis the performance across different types of users.

\textbf{Predictive Performance on RMSE.}

Table \ref{Res_RMSE} shows the predictive performance in RMSE for duration time over various deadlines. We show the predictive performance over deadlines from 30 days in the future to 90 days in the future. It can be seen from the table \ref{Res_RMSE} that our proposed models MHM, including MHMe and MHMl, outperform other methods on all duration deadline trials. Note that it is the introduction of price that make MHM have comparable performance over other methods in different deadline ranges. The effect of price factor in duration time prediction can be illustrated by the marginal improvement of HMH models to the CC baseline.

However, the structure discrepancy of hazard function does not lead to a nontrival changes in duration time prediction. It is shown in the comparisons among MHM models, i.e., the difference in performance of linear structure MHMl and the exponential structure MHMe is barely negligible. Similar situation can be seen in the comparison model, where the CC model is close to IB model in predicting performance. We also note that among all comparison methods, HP has the best results in prediction. It might due to the different approach on collected information. Note that HP captures the self-exciting information while others captures the neighbors influence.

\textbf{Predictive Performance on log-likelihood.}

In addition to RMSE, we also evaluate our models on test log-likelihood. Figure \ref{Res_TLL} shows the predictive performance on log-likelihood of test set among various datasets. Once again, our proposed models outperform other methods on all datasets. It is worth mentioning that, the Online Retail dataset is much denser than any dataset comes from Amazon. The collected influence thus plays a more important role in this case, which can be seem from the last subplot. Methods that include collected influence (HP, CC and IB) outperform other that dose not (PP). It can also be reflected that structure (linear structure vs exponential structure) of the hazard function does not effect much on the likelihood results.  

\textbf{Predictive Performance across users.}

\begin{figure*}
\centering
\includegraphics[height=1.85in, width=6in]{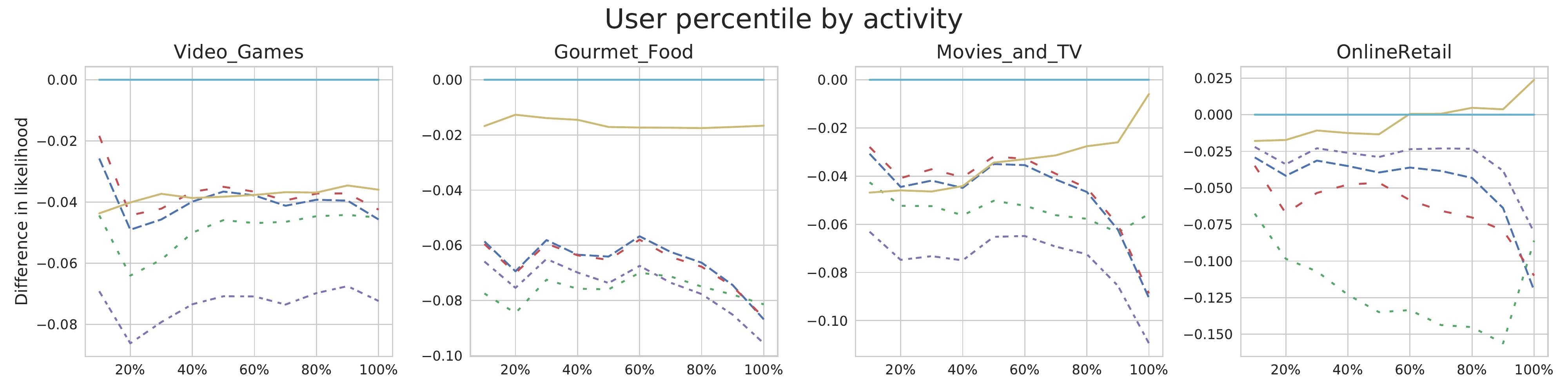}
\includegraphics[height=2.03in, width=6in]{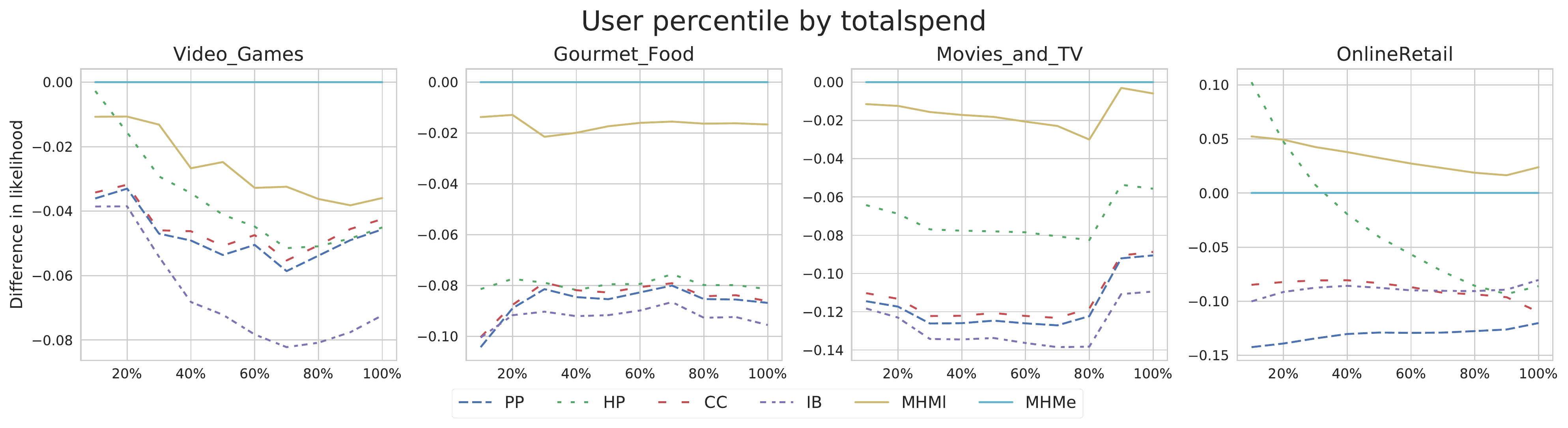}
\caption{Predictive performance across users. The top plots show the relative performance of methods across different types of users with respect to user activity $\mathcal{N}^{t}_u$. And the bottom plots show the relative performance across different types of users with respect to user total spend $\mathcal{P}_u$.}
\label{Res_User}
\end{figure*}

We also study predictive performance with respect to user activity $\mathcal{N}^{t}_u$ and user total spend $\mathcal{P}_u$ to investigate the difference in predicting across users of different types. Figure \ref{Res_User} shows the test log-likelihood results difference to our proposed MHM models. We measure the relative performance of methods across different types of users to MHM \cite{gopalan2015scalable}. In figure \ref{Res_User}, the percentile in the x-axis denotes bottom percentile of active user and most spending user respectively. For example, a 50\% in the bottom plots denote users from spending the least to middle spending. A 100\% means all users. The y-axis denotes the relative test log-likelihood comparison to MHM. A -0.05 suggests the method is 5\% worst than MHM in likelihood. From the figure, we can see that MHM outperform all other methods expect the least spending users (less than 30\%) from Online Retail but the MHM still outperforms its rival with the increase cover of users. In addition, the irregular of the relative performance curve of others comparing to MHMe is probably due to the different user activities distribution and user spending distribution across various datasets. Nevertheless, our proposed models MHM has overall the best performance across different types of users.

\textbf{Predictive Performance across purchases.}

\begin{figure*}
\centering
\includegraphics[height=5.1in, width=6in]{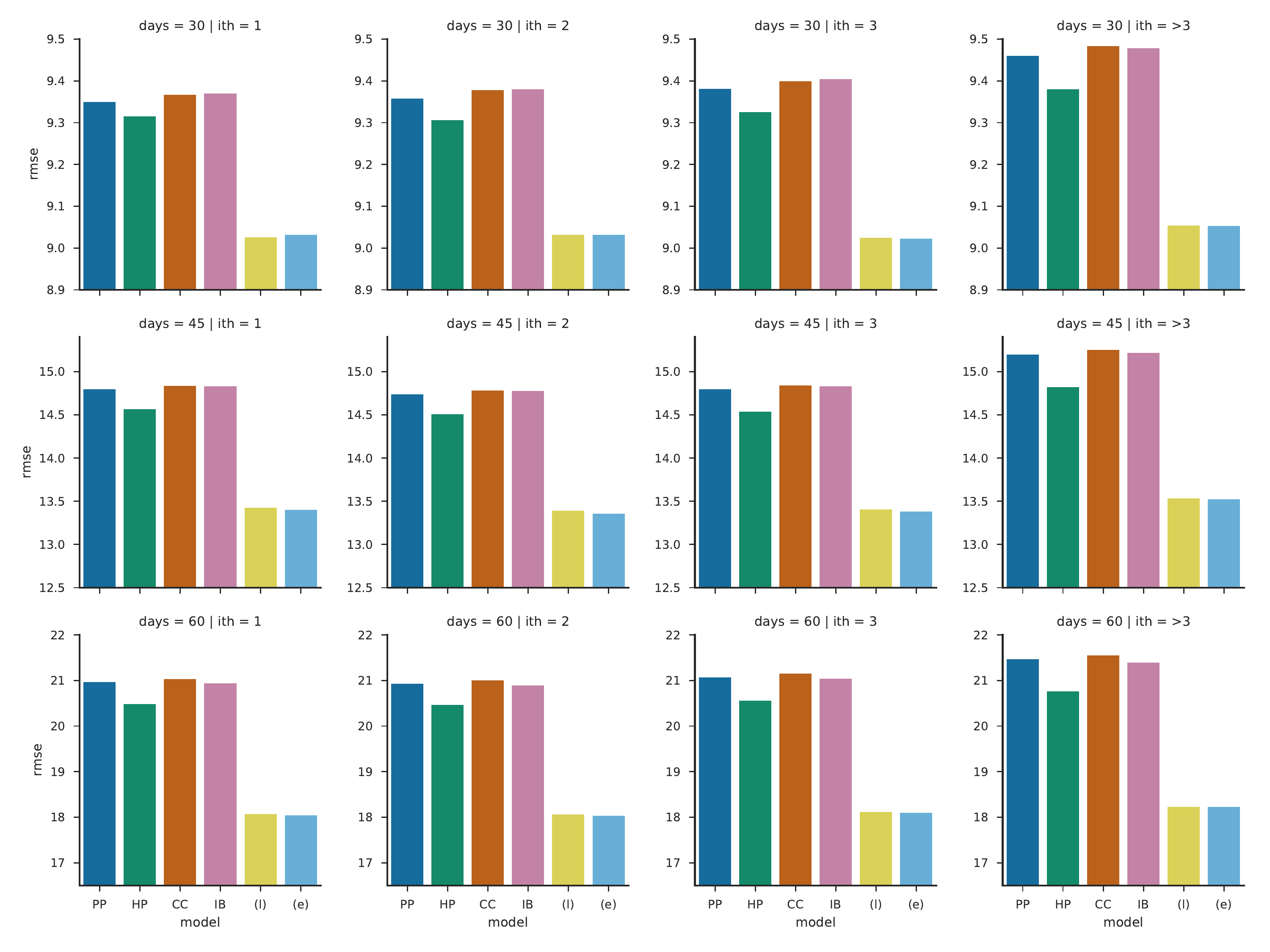}
\caption{Predictive Performance across purchases in test dataset. This figure shows the RMSE prediction results of purchase events for each user in the testing set with respect to their happening order of dataset Video Games. For example, days 60 and No. 2 denotes the prediction results of the second purchase within the next 60 days for each user.}
\label{Res_Ord}
\end{figure*}

we want to know if the prediction performance w.r.t. rmse increase with the order of purchase or if the model is immune to the order of purchase. This might reflect the confidence of the model for giving continuous predictions.
We divide the test dataset into different groups by the order of purchase for each user. For instance, a group might include the second purchase of all users within the next 60 days. As it is shown in figure \ref{Res_Ord}, each subplot is a prediction results of one group. Due to space limitation, we only able to show the resutls of two datasets. The proposed MHM models contrast distinguishable with other comparison models. And as the order of purchases increase, the MHM models keep stable performance on RMSE while other models have a slightly drop of performance specially when the order is greater than three. It is also worth mentioning that the MHM models marginally outperform its comparisons in larger prediction time period, which might indicate that price factor is more of a long-term impact.

\subsection{Posterior of spending}

\begin{figure}
\centering
\includegraphics[height=2.7in, width=3.1in]{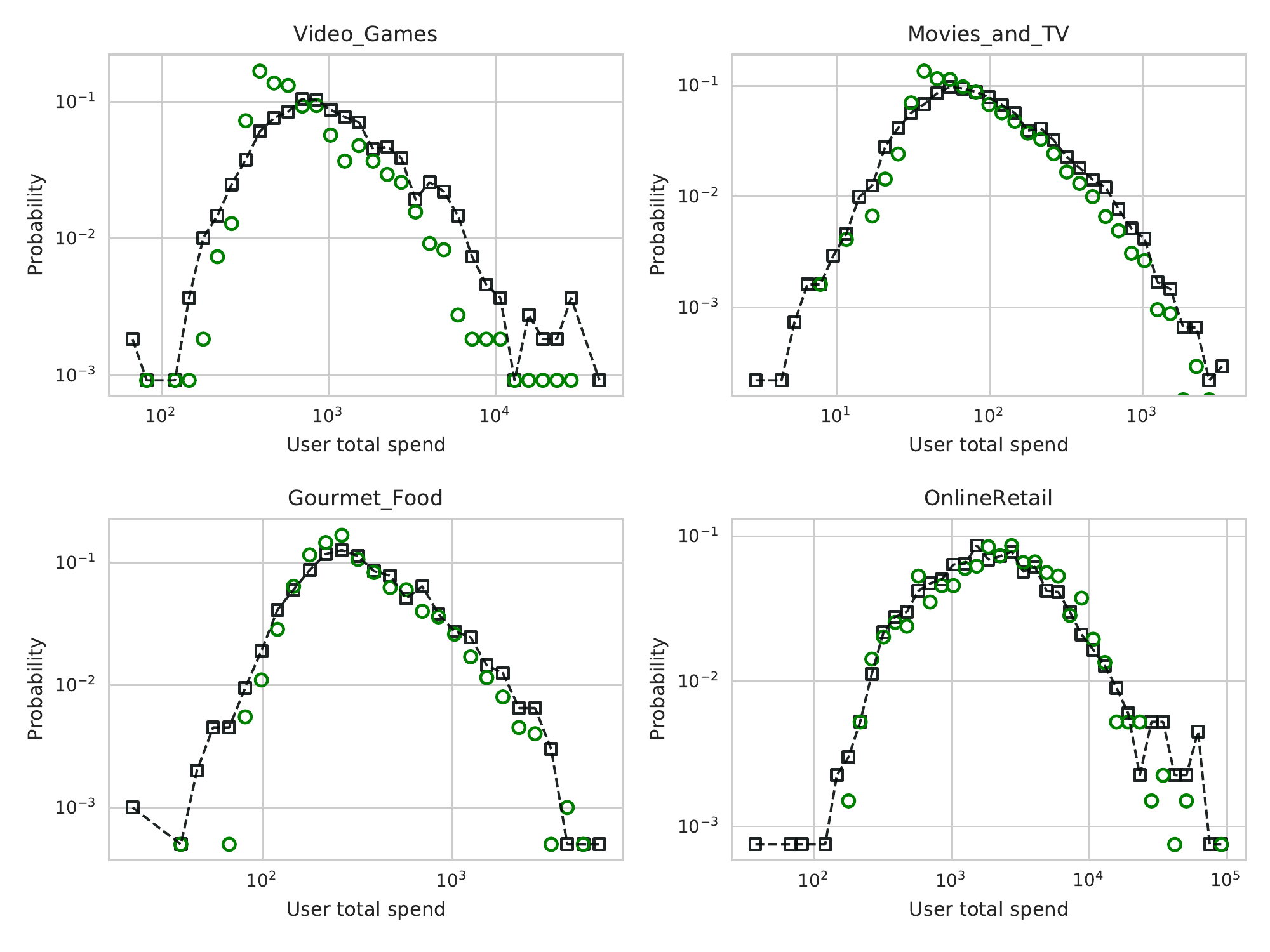}
\caption{Posterior samples on test dataset with respect to user total spend. It shows the samples on two datasets. The data scatter-dash lines with square shape denote the real spend distribution. The green circle scatters represent the samples from posterior.}
\label{Res_Samples}
\end{figure}

To investigate whether our price perception model (MHMe) can capture the price information from the dataset along with return time prediction, we draw samples from the spending posterior distribution of users and compare to the real user spending data. Figure \ref{Res_Samples} shows plots of posterior distribution comparison on each dataset. The real user total spend distribution is shown as black scatter-dash line in both plots and the predictive posterior distribution is shown as green scatter accordingly. It can be seen that both predictive distributions on two types of source match the real one appropriately, especially on the middle part. However, we also denote that in high spending area (for example, where total spend is over 10k on Video Games) the model is tend to underestimate the popularity due to the small amount of purchases with such high cost in the history purchasing data. Likewise, in low spending area the model is tend to overestimate the popularity might also due to the lack of samples in real situation.

\section{Conclusions}
We proposed a novel hazard based method that mixed price and social factor in traditional hazard function to model the return time of user purchasing and incorporate price influence on user decision making, whereupon the proposed MHM model can recommend the right item to a user considering its spendable resources when making a purchase. Moreover, with the Cox's hazard model characteristic, the MHM model can easily consider influence from neighbors or self-exciting. However, we found  the difference in duration time predictive performance of structure variance for the hazard function is quite limited. Experiments on two real world datasets demonstrate the improvement of the proposed MHM method over several comparison methods on both return time prediction  and test log-likelihood. Furthermore, our method also can outperform all comparison methods upon different types of user with respect to its activity and total spend and different purchases in testing set.

An interesting topic for future work is to utilize black box model like deep learning methods to construct an implicit hazard function since the prescribed  hazard function might limit its adaptability on real complex dataset. Another interesting topic is to develop new embedding algorithms to find high fidelity representations for events to feed the Cox's proportional hazard model. 


\bibliographystyle{ACM-Reference-Format}
\bibliography{acmart} 

\end{document}